\documentclass[conference]{IEEEtran}

\usepackage{ifpdf}
\usepackage{authblk}

\usepackage{cite}

\ifCLASSINFOpdf
  \usepackage[pdftex]{graphicx}
\else
  \usepackage[dvips]{graphicx}
\fi

\usepackage[cmex10]{amsmath}
\usepackage{amsfonts}
\usepackage{amssymb}

\usepackage{algorithmic}

\usepackage{array}

\usepackage{fixltx2e}

\makeatletter
\def\argmin{\mathop{\operator@font argmin}}
\def\argmax{\mathop{\operator@font argmax}}
\makeatother

\def\bm#1{\mbox{\boldmath $#1$}}

\newcommand{\bea}{\begin{array}}
\newcommand{\ena}{\end{array}}
\newcommand{\beq}{\begin{equation}}
\newcommand{\enq}{\end{equation}}

\newcommand{\beqa}{\begin{eqnarray}}
\newcommand{\enqa}{\end{eqnarray}}

\newcommand{\beqan}{\begin{eqnarray*}}
\newcommand{\enqan}{\end{eqnarray*}}

\newcommand{\AL}{\begin{enumerate}}
\newcommand{\ALE}{\end{enumerate}}

\def\addots{\mathinner{
    \mkern1mu\raise0pt\vbox{\kern7pt\hbox{.}}
    \mkern2mu\raise4pt\hbox{.}
    \mkern2mu\raise7pt\hbox{.}
    \mkern1mu}}

\def\sddots{\mathinner{
    \mkern.8mu\raise7pt\hbox{.}
    \mkern.8mu\raise4pt\hbox{.}
    \mkern.8mu\raise0pt\vbox{\kern7pt\hbox{.}}
    \mkern1mu}}

\def\saddots{\mathinner{
    \mkern.2mu\raise0pt\vbox{\kern7pt\hbox{.}}
    \mkern.2mu\raise4pt\hbox{.}
    \mkern.2mu\raise7pt\hbox{.}
    \mkern1mu}}

\makeatletter
\def\setboxz@h{\setbox\z@\hbox}
\def\wdz@{\wd\z@}
\def\boxz@{\box\z@}
\def\underset#1#2{\binrel@{#2}%
  \binrel@@{\mathop{\kern\z@#2}\limits_{#1}}}
\def\binrel@#1{\begingroup
  \setboxz@h{\thinmuskip0mu
    \medmuskip\m@ne mu\thickmuskip\@ne mu
    \setbox\tw@\hbox{$#1\m@th$}\kern-\wd\tw@
    ${}#1{}\m@th$}%
  \edef\@tempa{\endgroup\let\noexpand\binrel@@
    \ifdim\wdz@<\z@ \mathbin
    \else\ifdim\wdz@>\z@ \mathrel
    \else \relax\fi\fi}%
  \@tempa
}
\let\binrel@@\relax%
\makeatother

\def\circone{
    {\ooalign{
   \smash{\hskip3.0pt\raise0pt\hbox{\small $1$}}\vphantom{}\crcr
   \hbox{$\bigcirc$}
	}}
	} 
\def\circtwo{
    {\ooalign{
   \smash{\hskip3.0pt\raise0pt\hbox{\small $2$}}\vphantom{}\crcr
   \hbox{$\bigcirc$}
	}}
	} 
\def\circthree{
    {\ooalign{
   \smash{\hskip3.0pt\raise0pt\hbox{\small $3$}}\vphantom{}\crcr
   \hbox{$\bigcirc$}
	}}
	}

\def\sqplus{\mathbin{
	{\ooalign{\hfil\raise.3ex\hbox{\scriptsize
	+}\hfil\crcr\mathhexbox274\crcr\mathhexbox275}}
	}} 
\def\sqminus{\mathbin{
	{\ooalign{\hfil\raise.3ex\hbox{\scriptsize
	--}\hfil\crcr\mathhexbox274\crcr\mathhexbox275}}
	}}

\def\IC{{
   \mathord{
      \hbox to 0em{
	 \hskip-4pt
         \ooalign{
	   \smash{\hskip1.9pt\raise2.6pt\hbox{$\scriptscriptstyle |$}}\crcr
	   \smash{\hbox{\rm\sf C}} }
	 \hidewidth}
      \phantom{\hbox{\rm\sf C}}
} }}
\def\IN{
    {\ooalign{
   \smash{\hskip2.2pt\raise1.5pt\hbox{$\scriptscriptstyle |$}}\vphantom{}\crcr
   \hbox{\sf N}
	}}
	} 
\def\IZ{
    {\ooalign{
   \smash{\hskip1.9pt\raise0pt\hbox{$\sf Z$}}\vphantom{}\crcr
   \hbox{\sf Z}
	}}
	} 
\def\IR{
    {\ooalign{
   \smash{\hskip2.2pt\raise1.5pt\hbox{$\scriptscriptstyle |$}}\vphantom{}\crcr
   \smash{\hskip2.2pt\raise3.3pt\hbox{$\scriptscriptstyle |$}}\vphantom{}\crcr
   \hbox{\sf R}
	}}
	} 

\def\boldgreek#1{{
      \ooalign{
	\smash{\hskip.3pt\raise.1pt\hbox{$#1$}}\vphantom{}\crcr
	\smash{\hskip.5pt\raise.2pt\hbox{$#1$}}\vphantom{}\crcr
	\hbox{$#1$}\vphantom{$#1$}}}}

\iffalse
    \usepackage{amsbsy}

\else

\fi

\newcommand{\SI}{\begin{indlist}}
\newcommand{\EI}{\end{indlist}}

\newcommand{\DL}{\begin{dashlist}}
\newcommand{\DLE}{\end{dashlist}}

\begin{document}
\title{A Soft Range Limited K-Nearest Neighbours \\ Algorithm for Indoor Localization Enhancement}
\author{Minh Tu Hoang, Yizhou Zhu, Brosnan Yuen, Tyler Reese, Xiaodai Dong, Tao Lu, \\ 
Robert Westendorp, and Michael Xie}
\maketitle

\begin{abstract}
\footnote{Corresponding authors: X. Dong and T. Lu.\\
M. T. Hoang, Y. Zhu, B. Yuen, T. Reese, X. Dong and T. Lu are with the
Department of Electrical and Computer Engineering, University of Victoria,
Victoria, BC, Canada (email: \{xdong, taolu\}@ece.uvic.ca).\\
R. Westendorp and M. Xie are with Fortinet Canada Inc., Burnaby, BC,
Canada}
This paper proposes a soft range limited K nearest neighbours (SRL-KNN) localization fingerprinting algorithm. The conventional KNN determines the neighbours of a user by calculating and ranking the fingerprint distance measured at the unknown user location and the reference locations in the database. Different from that method, SRL-KNN scales the fingerprint distance by a range factor related to the physical distance between the user's previous position and the reference location in the database to reduce the spatial ambiguity in localization. Although utilizing the prior locations, SRL-KNN does not require knowledge of the exact moving speed and direction of the user. Moreover, to take into account of the temporal fluctuations of the received signal strength indicator (RSSI), RSSI histogram is incorporated into the distance calculation. Actual on-site experiments demonstrate that the new algorithm achieves an average localization error of $0.66$ m with $80\%$ of the errors under $0.89$ m, which outperforms conventional KNN algorithms by $45\%$ under the same test environment.  

\end{abstract}

%

\section{Introduction} \label{sec:intro}
The demand for accurate localization under indoor environments has increased dramatically in recent years with a large variety of the applications such as guidance, rescue operation, virtual reality game, etc \cite{Connolly2013 , Xu2016, C.Gentile2013}. For example, indoor positioning can help to guide customers in a shopping mall towards store, food court, etc., or passengers in an airport to the right terminal. In a museum, accurate indoor localization can transform a customer's phone into a virtual guide to give them contextual information based on his/her location. In this paper, the main application is to locate a walking human using the WiFi signals of the carried smartphone with an acceptable accuracy around a few feet.                           
In general, WiFi indoor localization methods can be grouped in two categories: one is signal propagation model based ranging, which utilizes received signal strength (RSS), the time of flight (TOF) and/or angle of arrival (AOA)\cite{HuiLiu2007} to estimate the location of the target; the other is fingerprinting based \cite{HuiLiu2007}, \cite{He2016}, which discriminates between locations by associating physically measurable properties as fingerprints or signatures for each discrete point. Due to the strong multipath effects, exact propagation model is difficult to obtain. Therefore, fingerprinting approach is more favourable for the WiFi based localization.\\
 \hspace*{0.2cm} Fingerprinting based WiFi localization can be realized by deterministic and probabilistic approaches \cite{He2016}. The former uses a similarity metric to differentiate the measured signal and the fingerprint data in the database before estimating the user's position as the closest fingerprint location in the signal space. Some typical examples of this approach are artificial neural network (ANN) \cite{Brunato2005}, \cite{Fang2008}, support vector machine (SVM) \cite{Brunato2005}, \cite{Shi2015} and K nearest neighbors (KNN) \cite{Bahl2000}, \cite{YaqinXie2016}, all of which require the collection of the fingerprints in the training phase to be compared with the measured signal in the testing phase for localization. Among these algorithms, ANN estimates location nonlinearly from the input by a chosen activation function and adjustable weightings \cite{Fang2008}. Despite its highest accuracy \cite{Laoudias2009}, \cite{C.Gentile2013}, this method is sophisticated in nature and requires extremely high computational complexity in the training phase \cite{Brunato2005}. In contrast, SVM is simpler than ANN \cite{Shi2015} but still relatively high in complexity. Compared to SVM and ANN, KNN has the lowest complexity while its accuracy is comparable to SVM \cite{Brunato2005}. On the other hand, the probabilistic algorithms are all based on statistical inference between the target signal measurement and stored fingerprints using Bayes rule \cite{Youssef2005}. Therefore, some probabilistic approaches assume the probability density function (PDF) of the RSSIs as empirical parametric distributions (e.g., Gaussian, double-peak Gaussian, lognormal \cite{L.Chen2013}, \cite{Xu2016}). This may not emulate the actual situation well \cite{QidengJiang2016}, leading to substantial localization errors. In order to achieve better performance, non-parametric methods \cite{Kushki2007, C.Figuera2009} make no assumption on the PDF of RSSI but require a large amount of data at each reference point, large storage and high computational resources to form the smooth and accurate PDF. 
Moreover, improvement of localization accuracy has been achieved by exploiting the measurements in previous time steps. For example, Kalman filter \cite{Au2013, Guvenc2003, Besada2007, Kushki2006} is used to estimate the most likely current location based on prior measurements, assuming a Gaussian noise and linear motion dynamics. In real scene, however, the assumption of Gaussian noise is not necessarily true \cite{Y.Chapre2013}, neither is the user's linear motion assumption a good approximation. A better motion model was proposed in \cite{Besada2007} with two Kalman filters, one for constant velocity case and the other one for a greater acceleration. The application of these two filters and switching in-between them increases the computational complexity significantly. In order to tackle the non-Gaussian and non-linear cases, extended Kalman filter \cite{Zhao2018} or particle filter \cite{Gustafsson2002, Evennou2005, Pak2017, Liu2017} can be applied. However, the major drawback of those filters is associated with high computational workloads and failure due to sample impoverishment \cite{Y.Chapre2013,Pak2017}. \\
\hspace*{0.5cm} This paper focuses on the study of KNN because of its low complexity suitable for practical use. In general, KNN computes the distance between the current WiFi RSSI fingerprint and the learned fingerprint in database to determine $K$ nearest neighbours. Different distance metrics such as Euclidean distance, Manhattan distance, and Mahalanobis distance can be used in KNN \cite{Xu2016}. Although being extensively investigated in literature, KNN still has the following open challenges:    
\begin{itemize}
\item[1] Spatial ambiguity \cite{CWu2017}: Some physically distant locations may have similar fingerprints or similar fingerprint distances compared with the current location.  This could mislead the KNN algorithms, leading to high localization errors.
\item[2] RSSI instability: Moving objects, constantly varying electromagnetic wave landscape in ambient environments, directionality of antenna and RF interference, etc., contribute to the wide fluctuation of WiFi signal \cite{Y.Chapre2013}. Therefore, the observed fingerprint of a location in the testing phase may not match that collected in the training phase. 
\item[3] RSSI short collecting time per location: Usually RSSI instability can be mitigated by taking the average of a large number of RSSI readings at one location. However, due to the mobile nature of the locating target, the RSSI sampling at each specific location in the testing phase is typically shorter than $2$ seconds. Within that duration, only a few number of RSSI readings can be collected. Consequently, the localization accuracy is severely impaired.   
\item[4] Heavy initial training phase: in order to construct the sufficient fingerprint map for accurate localization, a large number of reference points are required \cite{Jun2018}, which is time-consuming and labor-intensive \cite{He2016}.  
\end{itemize}    
To address the first three challenges, this paper incorporates the information of a user's previous position to KNN. Since the moving speed of the user in an indoor environment is bounded, the proposed soft range limited $K$ nearest neighbours (SRL-KNN) algorithm applies a penalty function based on the physical distance between the reference point and the anchor point (user's previous position) when calculating the fingerprint distance. As a result, the spatial ambiguity problem is significantly reduced. In contrast to other approaches such as Kalman filters that also exploit the measurements from previous time steps \cite{Au2013, Guvenc2003, Besada2007, Kushki2006}, our SRL-KNN method is much simpler and does not require the assumption of Gaussian noise distribution or linear motion. In addition, this paper proposes to use histogram and the combination of multiple fingerprints such as the mean, the difference of RSSI, the ranks of the AP RSSIs to tackle the RSSI instability and improve the localization accuracy. Actual on-site experiments show that our proposed algorithms can work well with the limited number ($1$ or $2$) of RSSI scans in each testing location (Section \ref{sec:Hist}).  

To reduce the work load in the initial training phase, crowdsourcing-based approaches have been proposed to replace the professional site survey with explicit and unprofessional user participation \cite{Wang2016,Zhao2018}. However, these methods are vulnerable to imperfect data, since the involved users are not always accustomed to the collecting systems \cite{He2016}. On the other hand, \cite{Liu2017, Jun2018} utilize relative RSSI differences among various access points (APs) called AP-sequence to reduce the number of required reference points (RPs). In \cite{Jun2018}, the area of interest is divided into a set of small regions based on AP-sequence and the user's location is estimated to be at the center of these regions. The main disadvantage of this approach is that the localization accuracy varies widely at different regions. Therefore, in addition to WiFi AP-sequence, \cite{Liu2017} also adopts the inertial-measurement unit (IMU) sensors and FM signal to refine the estimated location. In our experiment, we address the training phase challenge with the support of an autonomous robot. Our $3$-wheel robot (Fig. \ref{fig:fin_combine}(a)) has multiple sensors including wheel odometer, an inertial measurement unit (IMU), a LIDAR, sonar sensors and a color and depth (RGB-D) camera. The robot can navigate to a target location to collect WiFi fingerprints automatically. The localization accuracy of the robot is $0.07$ m $\pm$ $0.02$ m. Therefore, the time consumption and degree of human involvement for fingerprinting map construction is significantly reduced. 

The rest of the paper is organized as follows. Section \ref{sec:related_work} introduces related works on KNN, followed by details of SRL-KNN in Section \ref{sec:system}. Section \ref{sec:sim_result} reports the experimental set-up and results for the performance evaluation. Finally, Section \ref{sec:conclude} concludes this paper.      

\section{Related Work} \label{sec:related_work}
The original research on KNN indoor localization dates back to 2000 when a group from Microsoft demonstrated RADAR \cite{Bahl2000}. In that work, the mean and standard deviation of RSSI from multiple base stations are collected in the training phase and the Euclidean distance is used in the testing phase to determine the user's position. There are $70$ reference points (RPs) with $2.8$ m distance spacing (grid size). Testing points are picked randomly among these reference points. The average accuracy of this system is around $3$ m with $75 \%$ of the localization errors are below $4.7$ m.         

A refinement of the above method is the weighted KNN (WKNN) proposed by Brunato \textit{et al.}\cite{Brunato2005}, which calculates the user's position by the weighted average of the RSSI distance between estimated nearest neighbours and the current measurement. The experiment is implemented with $207$ reference locations, $50$ testing locations and a grid size of $1.7$ m. The accuracy of WKNN is $3.1 \pm 0.1$ m and $75\%$ of the localization errors are below $3.9$ m.          

To accommodate device heterogeneity, Zou \textit{et al.}\cite{Zou2017} proposed signal tendency index - weighted KNN (STI-WKNN) by adopting the similarity index STI between RSSI curve shapes to improve the localization accuracy. The raw RSSI signal is first transformed to a normalized object based on procrustes analysis (PA) method \cite{Gower1975}. Then signal tendency index is computed according to Euclidean distances between real time PA object and those stored in the fingerprint database. The final location will be determined by weighting among $K$ nearest neighbours that provide the smallest STI. Their experiment shows that STI-WKNN improves the localization accuracy by $23.95\%$ over the original WKNN across heterogeneous mobile devices.       

In a following research, Shin \textit{et al.}\cite{Shin2012} proposed to dynamically change the number of nearest neighbours $K$. Firstly, the RSSI Euclidean distance $D_{i}$ of each reference point $i$ is computed and $N$ numbers of which smaller than a threshold $T$ are picked. In a second step, the average of the selected $D_{i}$ is calculated to obtain a value $E$ and $K$ neighbours that satisfy $D_{i}<E$ are chosen. In general, this method only provides slightly lower average localization error than the classical KNN in RADAR \cite{Bahl2000}, except in the corridor where the testing scene is de facto one-dimensional. 
  
Taking into account the limited movement capabilities of a mobile user in an indoor environment, some researchers tried to utilize the information from the previous locations to improve the accuracy of KNN. In \cite{Khodayari2010}, Khodayari \textit{et al.} predicted the next probable location of the user by determining the speed and movement direction based on his/her last two recorded locations. Then, this prediction will  be considered only when the localization result of WKNN \cite{Brunato2005} is substantially deviated from the prior location. The underline assumption is that users moving at both constant speed and direction, which is not the case in many practical scenes. In \cite{Altintas2012}, Altintas \textit{et al.} added a short term memory which stores the recent signal strength observations as the historical data. In the testing phase, the current RSSI readings and all historical RSSI readings in the memory are added and taken the average. This helps to eliminate the unexpected signal strength readings due to the reflection, diffraction and scattering of the radio waves. However, this method is valid only when the variation of RSSI between the current and previous positions is small, which is not always true.  
 
In order to improve the localization stability, Xie \textit{et al.}\cite{YaqinXie2016} used Spearman distance based on the RSSI ranking between APs. According to \cite{YaqinXie2016}, although the absolute RSSI readings of a set of APs in a fixed location might be quite different, their rankings are more likely to remain the same, making it feasible to form a stable fingerprint. The drawback is that this algorithm is limited by the number of APs available. In the simulation of \cite{YaqinXie2016}, there are $400$ reference locations but only $4$ APs which can provide a maximum of $4!\, = \, 24$ ranking fingerprints. Consequently, many different locations have the same fingerprints, leading to localization errors in the testing phase.
 
In general, all of the above methods provide acceptable accuracy within around twice the distance between two consecutive reference points (grid size), but the problems of KNN algorithm mentioned in Section \ref{sec:intro} are not effectively solved. For example, previous KNN research have not sufficiently investigated the inadequate sampling of RSSI due to the user's movement, i.e., only $1$ or $2$ RSSI readings are available in each testing location. Obviously this ignores a very important factor and affects the localization accuracy. In addition, in the methods that use historical data, the assumption that users moving in constant speed and direction is unrealistic in many scenes. Therefore, a new KNN algorithm, which addresses the aforementioned problems of KNN, is proposed.
\section{System Model} \label{sec:system}
\subsection{Localization Scene}
\begin{figure}[!t]
\centering
\includegraphics[width=0.48\textwidth]{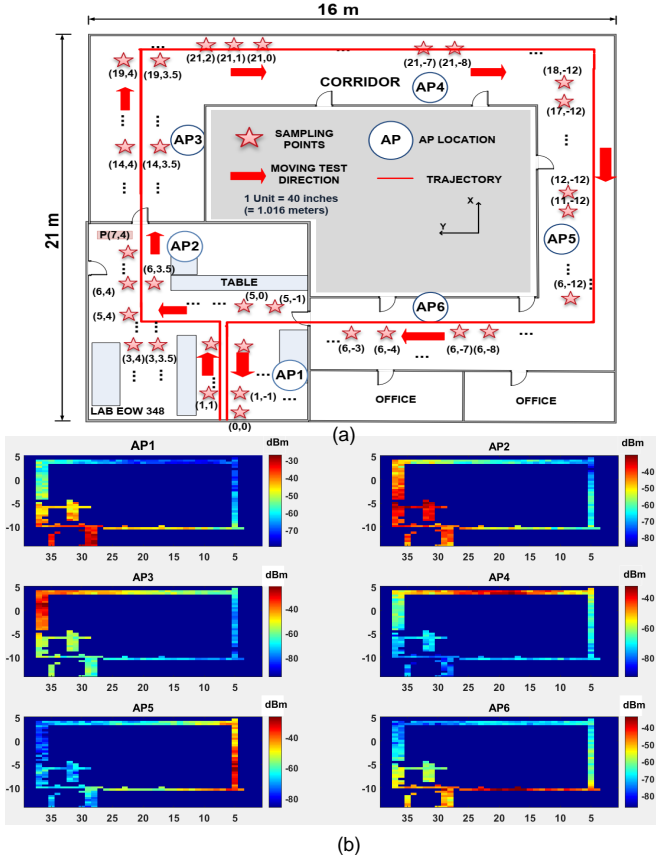}
\caption{(a) Floor map of the test site. The solid red line is the mobile user's walking trajectory with red arrows pointing toward walking direction. (b) Heat map of the RSSI strength from $6$ APs used in our localization scheme.}
\label{fig:floor_map}
\end{figure}

The fingerprinting localization system is generally divided into two phases: a training phase (offline phase) and a testing phase (online phase). In the training phase, features of the WiFi signals at each predefined reference point (RP) location, are collected and stored to a database. Those features typically include the mean and standard deviation of RSSI, the RSSI ratio between a pair of APs, the ranks of the APs, etc \cite{Xu2016}. They individually or collectively form fingerprints at each RP. Here, we assume the area of interest has $P$ APs and $M$ RPs. For each RP $i$ at its physical location $\bm{l}_{i}(x_{i},y_{i})$, a corresponding fingerprint vector is denoted as $\bm{f}_{i}=\{F^{i}_{1}, F^{i}_{2},...,F^{i}_{N}\}$, where $N$ is the number of available features and $F^{i}_{j}(1 \leq j \leq N)$ is the $j$-th feature at point $i$. In the testing phase, each unknown location of the user, denoted as a testing point, is determined by the localization algorithm. During the training phase, multiple RSSI scans ($S_{1}$ scans) can be obtained at each location, and hence a set of RSSI values correspond to one RP while in the testing phase, only a small number of RSSI readings ($S_{2}$ scans), e.g., $S_{2}=1$ or $S_{2}=2$, is available for the fingerprint matching. Fig. \ref{fig:floor_map}(a) illustrates our localization scheme with $6$ APs, $365$ RPs and $175$ testing points. Fig. \ref{fig:floor_map}(b) shows the heat map of $6$ APs, where we represent signal strength by color. Clearly, the signals from $6$ APs already cover the whole targeting area including $1$ room and $3$ corridors.                
\subsection{The classical KNN algorithm}
The fingerprint distance between the unknown current point $\bm{l}$ and each reference point $i$ in database is first calculated as follows    
\begin{equation} \label{eq:NormalEu}
D^{i}_{l}= \sqrt{\sum_{j=1}^{N} (F_{j}-F^{i}_{j})^2}
\end{equation}
where $F_{j}$ is the $j$-th fingerprint feature at the unknown location, $N$ is the number of available fingerprints. Then $K$ locations with the minimum distances are chosen as the $K$ nearest neighbours. Finally, the position $\bm{l}$ of the user is determined by taking the average of all those $K$ neighbours' locations.   
\subsection{Proposed Soft Range Limited KNN (SRL-KNN) method}

\subsubsection{SRL-KNN algorithm}
This paper proposes to leverage the information of the user's previous position, as the moving speed of a user is limited and one cannot instantaneously move to an unrealistic distant position from the prior one during the consecutive measurements. In a simple form, a circle can be drawn around the previous location to limit the nearest neighbour search space to within the circle, whose radius is determined by the user moving speed and time duration between two consecutive measurements. Instead of using that hard range limit, we here devise a novel soft range limiting factor to the fingerprint distance calculation where the locations near the user's previous position are given higher likelihood to become one of $K$ nearest neighbour candidates. To achieve that, we modify the Euclidean distance in (\ref{eq:NormalEu}) as follows
\begin{equation} \label{eq:PenaltyEu}
{\bar{D}_{l}}^{i}= \frac{W^{i}_{l} \times D^{i}_{l}}{\sum_{i=1}^{M} W^{i}_{l}}
\end{equation}
\begin{equation} \label{eq:weight} 
W^{i}_{l} = \exp (\frac{(x_{i}-x_{pre})^{2}+(y_{i}-y_{pre})^{2}}{4\sigma^{2}})
\end{equation}
where $W^{i}_{l}$ is the penalty function for the location $i$, $M$ is the total number of RPs in the database, $(x_{pre},y_{pre})$ is the most recent previous location of the user, $\sigma$ is the maximum distance which the user can move in a consecutive sampling time interval $\Delta{t}$. For example, people tend to walk in indoor environments at a speed from $0.4$ m/s to $2$ m/s \cite{Browning2006}, \cite{Email2007} (maximum speed $v_{max} = 2$ m/s) and the user location will be updated every $1$ second (consecutive sampling time interval $\Delta{t} = 1$ s). Therefore, $\sigma \, = \,v_{max}\,\,\Delta{t} \,=\, 2$ m. As shown in Fig. \ref{fig:fin_combine}(c), the penalty function has the form of a Gaussian distribution with the mean being the previous location and the standard deviation being $\sigma$. Note that the prior position is only used here to form the soft range limit scaling factor as shown in (\ref{eq:weight}), unlike in Kalman filter approaches which directly include the history position in the current location calculation. Moreover, our formulation only assumes a maximum moving speed, but does not require knowledge of the exact moving speed and direction of the user. The user's location \bm{l} is determined through a weighted average of $K$ nearest neighbours $\bm{l}_{j}$ as follows     
\begin{equation} \label{eq:wknn}
\bm{l} = \frac{\sum_{j=1}^{K} \frac{\bm{l}_{j}}{\bar{D}_{l}^j}}{\sum_{j=1}^{K} \frac{1}{{\bar{D}_{l}}^{j}}} 
\end{equation}
where ${\bar{D}_{l}}^{j}$ is the modified Euclidean fingerprint distance which was presented in (\ref{eq:PenaltyEu}). 

 \subsubsection{Fingerprint combination} \label{sec:Fingerprint}
In the WiFi fingerprinting method, the more stable the fingerprint is, the better the localization accuracy will be. However, the RSSI collected by a client device often experiences substantial fluctuations due to dynamically changing environments such as human blocking and movements, interference from other equipment and devices, receiver antenna orientation, etc., \cite{YogitaChapre2013}, \cite{Kaemarungsi2012}. Therefore, this papers proposes to use the combination of a set of different fingerprints to ensure sufficient stability and distinctive values in each location. The most common fingerprint used is the mean of RSSI \cite{Brunato2005}, \cite{Bahl2000} which fluctuates significantly due to the previously mentioned effects. In contrast, one of the more reliable fingerprints is the mean difference of RSSI between a pair of APs. In \cite{F.Dong2009}, Dong \textit{et al.} used two devices, i.e., a laptop and a smart phone to collect RSSI in a fixed location. They observed that although the individual RSSI readings of these devices fluctuate significantly, the mean differences of RSSI between pairs of APs are more stable. Therefore, the mean difference of RSSI can be used to address the received signal strength offset problem between different mobile devices. In addition, the rank fingerprints described in \cite{YaqinXie2016} can also be used as an additional fingerprint if there are enough number of APs available. Recently, Tian \textit{et al.}\cite{Tian2018} utilize a new fingerprint named temporal correlation of the RSSI to improve the location estimation accuracy. However, in order to get the stable RSSI temporal correlation, a sufficient number of RSSI readings in each testing location is required, which is not feasible in our test cases.     
\begin{figure}[!t]
\centering
\includegraphics[width=0.52\textwidth]{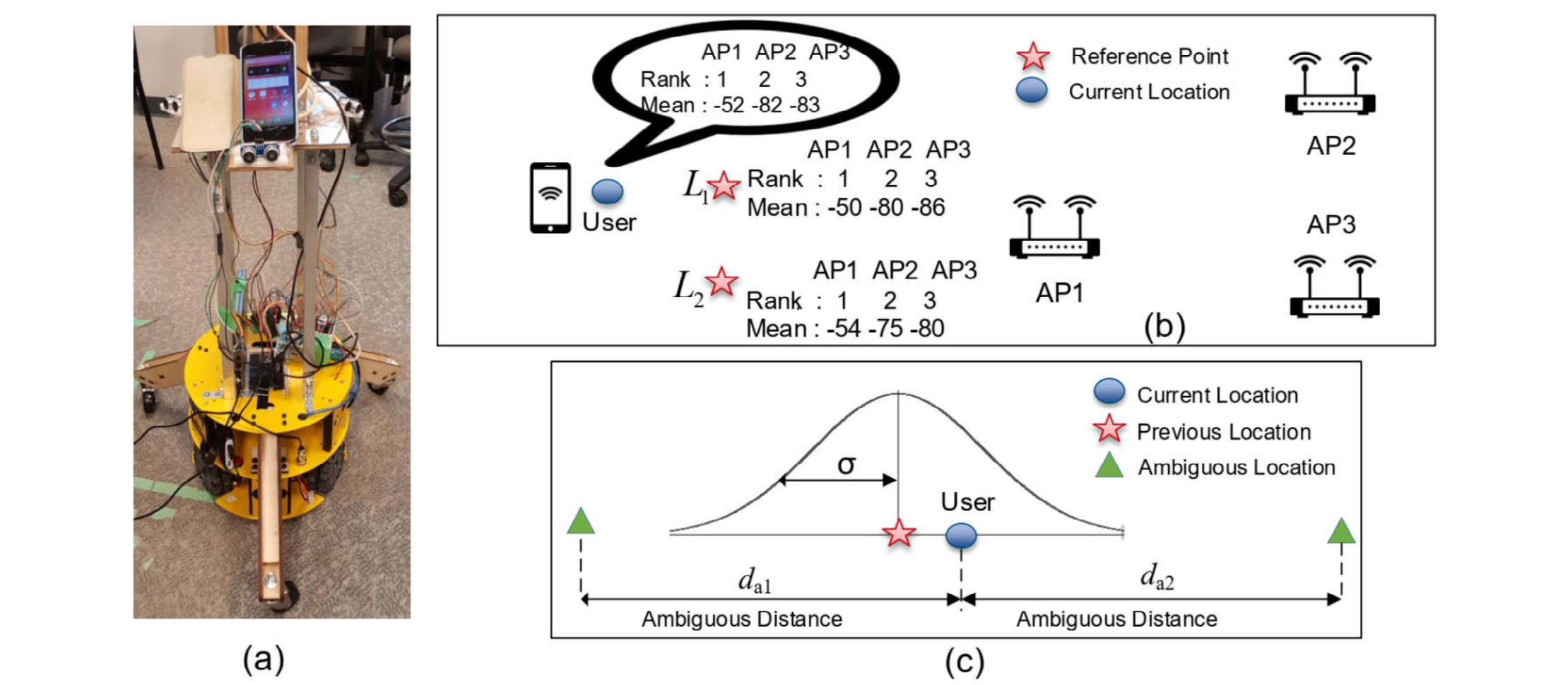}
\caption{(a) 3-wheel robot. (b) Fingerprint combination illustration. (c) Penalty function illustration.}
\label{fig:fin_combine}
\end{figure}              
In our experiment, we first utilize some fingerprint types such as the RSSI differences and/or the AP rank to get $n$ nearest neighbours RPs according to the shortest distance computed from (\ref{eq:PenaltyEu}). Within the chosen nearest neighbours, we then refine our selection to $K$ ($K<n$) nearest neighbours by using the mean of RSSI as the fingerprint. For example, Fig. \ref{fig:fin_combine}(b) illustrates the scenario where we have a user trying to locate his location with the information of both the mean of RSSI and the rankings from $3$ different APs. By using the rank fingerprints, two neighbours $L_1$ and $L_2$ are chosen based on the minimum fingerprint Euclidean distances. However, these points have the same rank fingerprints so we need to use the mean of RSSIs as the additional information to determine which point is the true neighbour of the user's location. With regard to mean fingerprints, neighbour $L_1$ that provides the smaller Euclidean distances is more likely the exact neighbour which we want to find.                    
\subsubsection{Histogram of RSSI} \label{sec:Hist}
As mentioned above, the raw RSSI readings at a location are unstable, fluctuating widely up to $10$ dB \cite{Y.Chapre2013}. Therefore, they may not represent well the feature of the RSSI at each location. In order to solve this problem, one may include the histogram of RSSI in the fingerprint distance calculation, which defines the probability of the original RSSI  reading of the $j$th AP falling into [$R_j-0.5$ dBm, $R_j+0.5$ dBm] at the reference location $i$ as follows \cite{LeiYang2015}
\begin{equation} \label{eq:histogram}
p_{R}^{i,j} = \frac{n_{R_j}^{i}}{n_{total}^{i,j}}
\end{equation}  
where $n_{total}^{i,j}$ is the total number of RSSI scans of the $j$th AP at location $i$, $n_{R_j}^{i}$ is the number of RSSI readings of the $j$th AP falling into the range between $R_j-0.5$ dBm and $R_j+0.5$ dBm ($R_{L}^{j} \, \leq R_j \leq R_{U}^{j} \,$), $R_{L}^{j}$ and $R_{U}^{j}$ are the minimum and maximum values of RSSI of $j$th AP respectively. Consequently, (\ref{eq:NormalEu}) can be modified as a weighted distance according to
\begin{equation} \label{eq:His_Euclidean}
D_{l,hist}^{i}= \sqrt{\sum_{j=1}^{N} \sum_{R_j=R_{L}^{j}}^{R_{U}^{j}} p_{R}^{i,j}(F_{j}-R_{j})^2}
\end{equation}   
and the final fingerprint distance is obtained as   
\begin{equation} \label{eq:PenaltyEuHist}
{\bar{D}_{l}}^{i}= \frac{W^{i}_{l} \times D_{l,hist}^{i}}{\sum_{i=1}^{M} W^{i}_{l}}
\end{equation}
\section{Experiment And Analysis} \label{sec:sim_result}
\subsection{Experimental Setup}
All experiments have been carried out on the third floor of Engineering Office Wing (EOW), University of Victoria, BC, Canada. The dimension of the area is $21$ m by $16$ m. It also has $3$ long corridors as shown in Fig. \ref{fig:floor_map}(a). The RSSI measurements were taken in $365$ pre-determined RPs. A mobile device (Google Nexus $4$ running Android $4.4$) mounted on a $3$-wheel robot (Fig. \ref{fig:fin_combine}(a)) was sent to target locations to collect fingerprints. The localization accuracy of the robot is $0.07$ m $\pm$ $0.02$ m. At each location, $100$ instantaneous RSSI measurements ($S_{1}=100$) were collected to a database. There are $6$ APs and $5$ of them provide $2$ distinct MAC address for $2.4$ GHz and $5$ GHz communication channels respectively. Equivalently, in every scan, $11$ RSSI readings from those $6$ APs can be collected. 

In the testing phase, we conducted both one-location test and trajectory test. In the one-location test, RSSI values at a fixed position were collected and the user's position was determined in every consecutive sampling time interval $\Delta{t}$. In the trajectory test, the robot carried a mobile device and moved along the direction as shown by the red solid line in Fig. \ref{fig:floor_map}(a). RSSI readings were collected continuously by the phone and were transmitted to a server in real time. The server analyzed the data to locate the user's position. The mean fingerprint in each location was determined by the average of $S_{1}$ RSSI readings for training and $S_{2}$ RSSI readings for testing. On the other hand, the mean difference of RSSI fingerprint for a test location was calculated by taking the average of $S_{1}$ ($S_{2}$) RSSI differences between a pair of APs.           

\subsection{One-Location Test}
\begin{table*}[!t]
\centering         
\caption{Average localization errors} \label{table:AverageErr} 
\begin{tabular}{l c c c c c c c} 
\hline           
\textbf{Method} & \textbf{SRL-KNN Mean} & \textbf{SRL-KNN Rank} & \textbf{SRL-KNN Histogram} & \textbf{SRL-KNN Mean and Rank} & \textbf{SRL-KNN Mean and RSSI Differences}\\ 
Average Error (m) & 0.81 $\pm$ 0.40 &  1.20 $\pm$ 0.96  &  0.66 $\pm$ 0.36 &  0.76 $\pm$ 0.51 & 0.71 $\pm$ 0.46\\
\textbf{Method} & \textbf{RADAR\cite{Bahl2000}} &  \textbf{STI-WKNN\cite{Zou2017}}  &  \textbf{Spearman Rank\cite{YaqinXie2016}} & \textbf{Kernel Method\cite{Kushki2007}} & \textbf{Kalman  Filter\cite{Au2013}}\\ 
Average Error (m) & 1.19 $\pm$ 0.86  &  1.09 $\pm$ 0.81 & 1.45 $\pm$ 1.14 &  1.07 $\pm$ 0.86 &  0.96 $\pm$ 0.48 \\
\hline         
\end{tabular} 
\end{table*}

\begin{figure}[!t]
\centering
\includegraphics[width=0.5\textwidth]{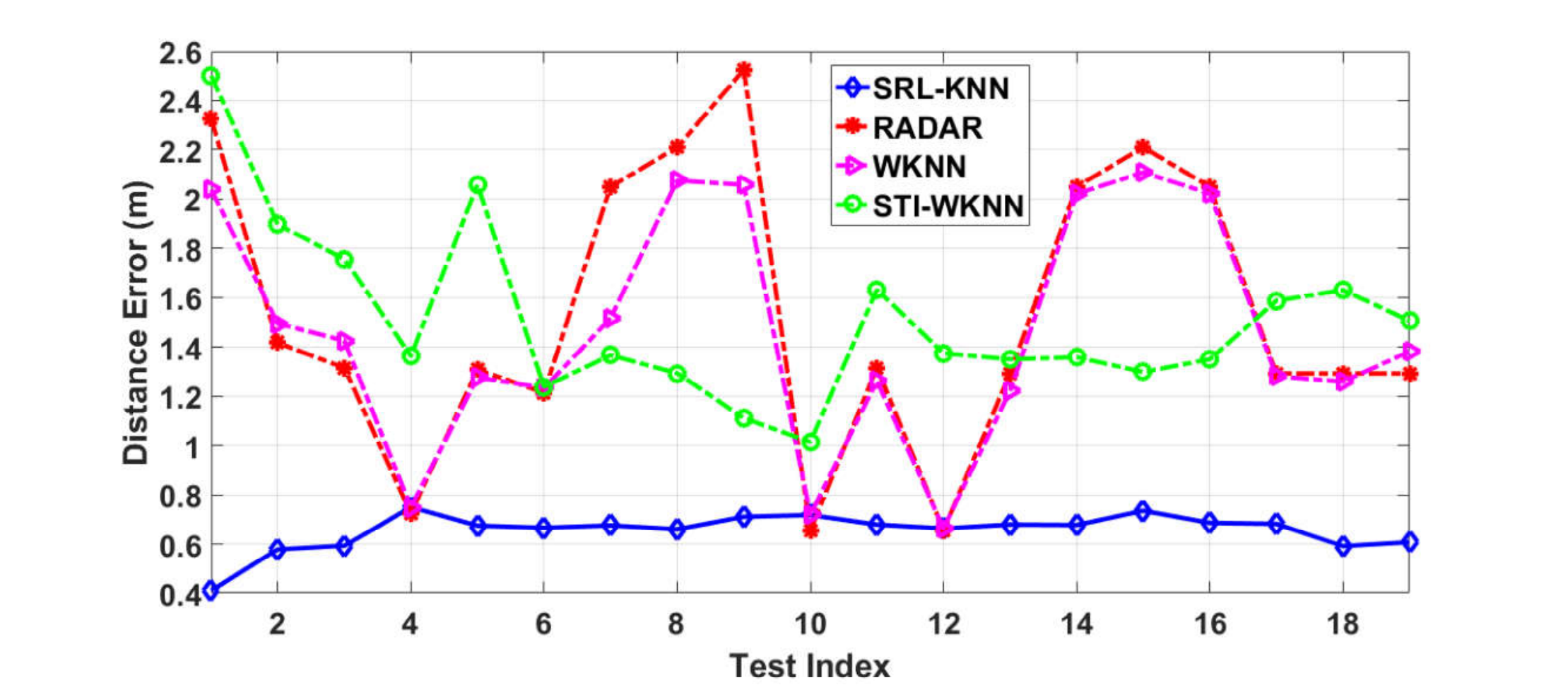}
\caption{Localization errors of one-location test.}
\label{fig:fixed_test}
\end{figure}
In this test, the mobile device was put on the location $P(7,4)$ as shown in Fig. \ref{fig:floor_map}(a). The experiment was conducted in busy hours when many students (up to $10$) used WiFi and moved around the lab. A maximum RSSI standard deviation of $5.5$ dB was recorded over $100$ consecutive RSSI readings. The large fluctuation of RSSI is due to the factors explained in Subsection. \ref{sec:Fingerprint}.  

Fig. \ref{fig:fixed_test} shows the comparison of the localization accuracy among the classical KNN fingerprinting algorithms in RADAR \cite{Bahl2000}, WKNN \cite{Brunato2005}, STI-WKNN \cite{Zou2017} and our proposed SRL-KNN algorithm. All algorithms use the mean of RSSI as the fingerprint, the consecutive sampling time interval $\Delta{t} = 1$ s and the number of nearest neighbours $K=3$. The user location is estimated based on $1$ RSSI scan ($S_{2}=1$) collected every $\Delta{t}$. Over all $19$ tests conducted at different time instants within one hour, the localization results of RADAR, WKNN and STI-WKNN fluctuate more than $1.7$ m from $0.70$ to over $2.40$ m, while SRL-KNN reports a much lower fluctuation with $0.3$ m from $0.40$ to $0.70$ m. The accuracy of SRL-KNN is $2$ times better than the other methods with average distance error being $0.60$ m compared with over $1.20$ m of the others.   
   
\subsection{Trajectory Test} \label{sec:Traj}
In this test, the robot moved along a pre-defined route as shown in Fig. \ref{fig:floor_map}(a) with an average speed around $0.6$ m/s. All the testing locations (total $175$ locations) along the trajectories are randomly picked. In this experiment, the maximum speed in our algorithm is set to $v_{max} \, = \, 2$ m/s, so the maximum distance which user can move is $\sigma \, = \,v_{max} \times \Delta{t} \,=\, 2$ m. The initial position of the user in these testing trajectories is assumed to be known. All the other parameters are the same as those in the one-location test. 
\begin{figure}[!t]
\centering
\includegraphics[width=0.5\textwidth]{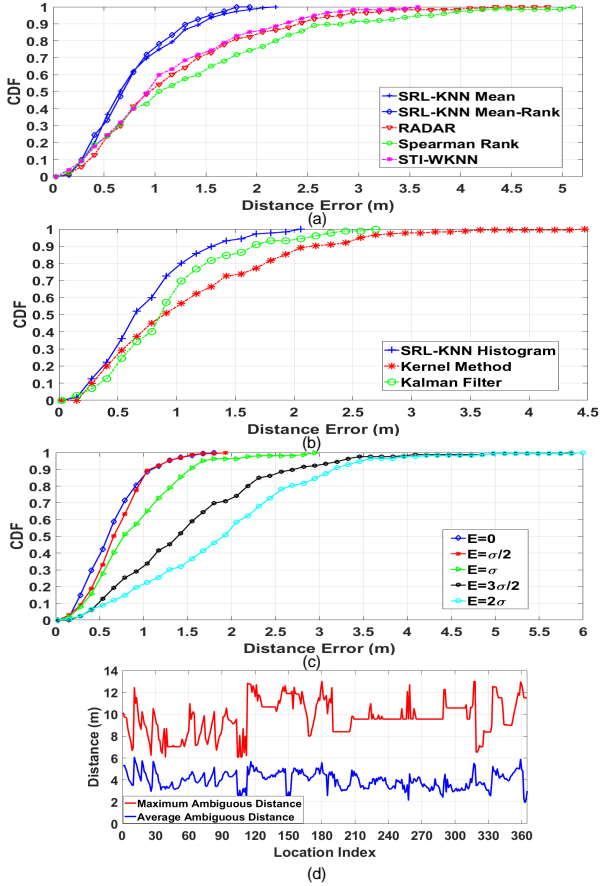}
\caption{(a) CDF of localization errors of SRL-KNN using mean and rank database and other KNN methods. (b) CDF of localization errors of SRL-KNN using histogram and other probabilistic methods. (c) CDF of localization errors of SRL-KNN using histogram in different error scenarios of historical data. (d) Maximum and average ambiguous distances of $365$ locations in the database}
\label{fig:CDF}
\end{figure}

\begin{figure}[!t]
\centering
\includegraphics[width=0.46\textwidth]{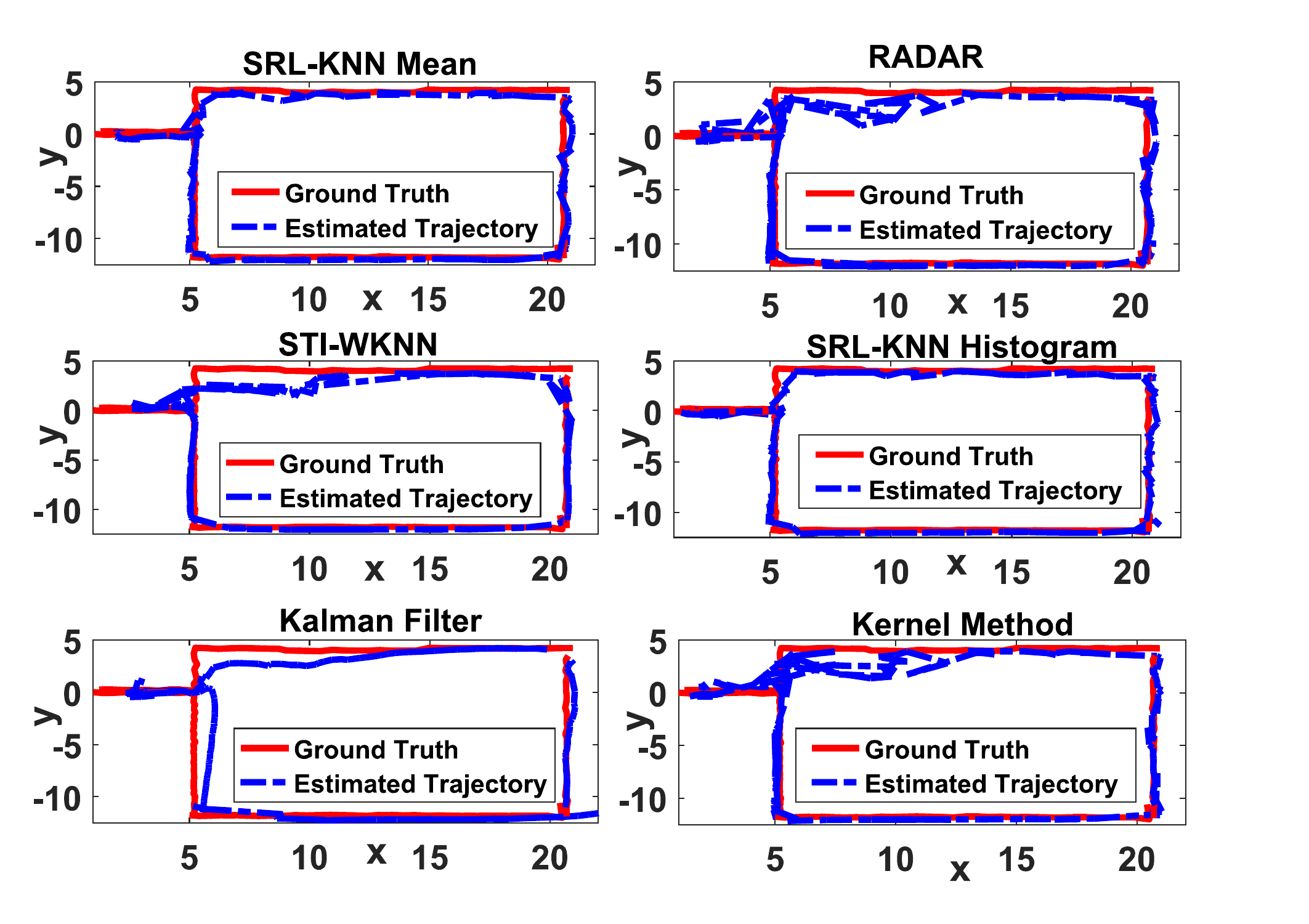}
\caption{Ground truth and estimated trajectories. Red line represents the trajectory ground truth. Blue lines are estimated trajectories}
\label{fig:GroundTruth}
\end{figure}
Fig. \ref{fig:CDF}(a) compares the cumulative distribution function (CDF) of localization errors between SRL-KNN and other KNN methods, i.e., RADAR \cite{Bahl2000}, Spearman rank distance \cite{YaqinXie2016}, STI-WKNN \cite{Zou2017}. Here, for comparison, we used both mean of RSSI, rank of APs as our fingerprints. Clearly, the SRL-KNN (blue line) outperforms the other methods in terms of positioning accuracy. Further analysis shows that due to larger RSSI fluctuations, the other methods may choose a wrong location with similar fingerprints as its nearest neighbours. Note that such location could be far from the actual location, leading to an extreme large error in the scale of the testing site dimension. As shown in Fig. \ref{fig:CDF}(a), a $4.80$ m maximum localization error is recorded for RADAR, $3.50$ m for STI-WKNN and the largest maximum localization error of over $5$ m for Spearman rank method. In contrast, SRL-KNN eliminates such error pattern, resulting in a much smaller maximum errors of $2.20$ m with the mean fingerprint. In particular, SRL-KNN using only mean fingerprint has $80\%$ of the location error within $1.20$ m while RADAR, STI-WKNN and Spearman rank distance are $1.80$ m, $1.80$ m and $2.30$ m respectively. To achieve higher accuracy, the combination of different fingerprint described in Subsection \ref{sec:Fingerprint} is used. In this article, we implemented two different fingerprint combinations: use the mean RSSI with the rank fingerprint and use the mean RSSI with RSSI difference between a pair of APs. In both cases, the rank or RSSI difference fingerprint is firstly utilized to get $n=7$ neighbours and then $K=3$ refined nearest neighbours are chosen based on the mean fingerprint. These two methods have the similar performance with the maximum error of around $1.80$ m and $80\%$ of the error is within $1$ m. 

We further implement the histogram based fingerprint distance described in Subsection \ref{sec:Fingerprint}. In the testing phase, the feature $F_{j}$ in (\ref{eq:His_Euclidean}) is obtained as the mean of all $S_{2}$ RSSI readings from an AP. In comparison  with the other probabilistic approaches including Kernel method \cite{Kushki2007}, Kalman  filter \cite{Au2013} in Fig. \ref{fig:CDF}(b), this approach clearly outperforms. Our method (plus markers) has a maximum error of $2.10$ m while the maximum errors of Kalman filter (circle markers) and Kernel method (star marker) are $2.70$ m and $4.50$ m, respectively. The $80 \%$ of the error in our histogram approach is $0.90$ m, following by $1.40$ m of Kalman filter and $1.90$ m of Kernel method. Fig. \ref{fig:GroundTruth} illustrates the ground truth and estimated trajectory using different methods. As clearly shown, both histogram and mean fingerprint SRL-KNN are the most accurate predictions. In addition, Table \ref{table:AverageErr} lists all the average localization errors. The best accuracy is $0.66$ m in the case of SRL-KNN using the RSSI histogram. Regarding the computational complexity, SRL-KNN has the similar complexity $O(MN)$ with the conventional KNN RADAR \cite{Bahl2000}, where $M$ is the number of RPs, and $N$ is the number of available features.  
  \begin{table}
\centering         
\caption{Average Localization Errors Of UJIIndoorLoc Database} \label{table:AverageErr2} 
\begin{tabular}{l c c c c c c} 
\hline           
 & \textbf{SRL-KNN Mean} & \textbf{RADAR \cite{Bahl2000}} & \textbf{STI-WKNN \cite{Zou2017}} \\ 
Building 0 (m) & 4.7 $\pm$ 2.7 &  7.9 $\pm$ 4.9 &  7.9 $\pm$ 5.2 \\
Building 1 (m) & 4.6 $\pm$ 3.8 &  8.2 $\pm$ 4.9 &  6.8 $\pm$ 6.1  \\
Building 2 (m) & 6.0 $\pm$ 4.5 &  8.2 $\pm$ 7.4 &  6.1 $\pm$ 3.7 \\
All buildings (m)& 5.0 $\pm$ 3.7 &  7.7 $\pm$ 6.0 &  7.0 $\pm$ 4.9 \\ 
\hline   
\end{tabular} 
\end{table}

\begin{figure}
\centering
\includegraphics[width=0.48\textwidth]{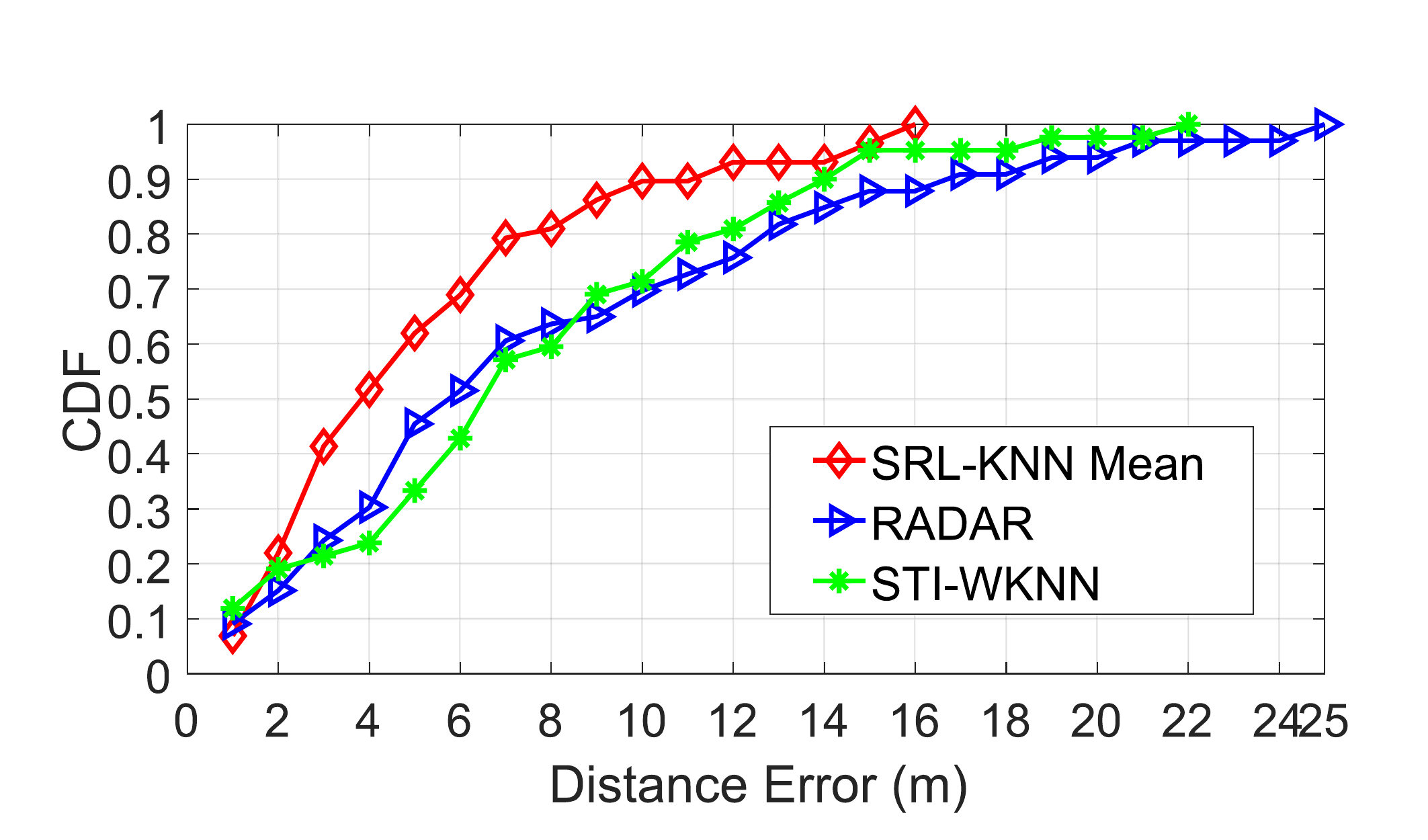}
\caption{CDF of localization errors of UJIIndoorLoc database for all $3$ buildings}
\label{fig:Ujdata}
\end{figure}   

Since SRL-KNN leverages the information of a user's previous position to estimate the current location, the performance of SRL-KNN depends on the accuracy of historical data. Note that all of our SRL-KNN results presented so far are based on the estimated imperfect history data. In order to look into the propagation error due to the imperfect prior location estimation, Fig. \ref{fig:CDF}(c) illustrates the localization errors of SRL-KNN using histogram based fingerprint distance with both the ideal and erroneous history data. Starting with the perfect historical coordinate $\bm{h}(x,y)$ for every location in the testing trajectories mentioned in Sec. \ref{sec:Traj}, an amount of error $E$ m is added to $\bm{h}$. The erroneous prior location $\bm{h^{\prime}}(x^{\prime}, y^{\prime})$ is obtained as: $ x^{\prime} = x + x_{e} \, , \, y^{\prime} = y + y_{e}$, where $x_e$ and $y_e$ are random variables that follow Gaussian distribution 
\[ x_e \sim \mathcal{N}(0,\sigma_{x_{e}}^{2}) \, ; \, y_e \sim \mathcal{N}(0,\sigma_{y_{e}}^{2}) \, ; \, \sqrt{\sigma_{x_{e}}^{2}+\sigma_{y_{e}}^{2}} = E \] 
Fig. \ref{fig:CDF}(c) shows the cases where $E$ is proportional to $\sigma=2$ m. Obviously, if the error $E$ of the history data is within $\sigma/2$ m, the localization accuracy is mostly similar to the ideal case, with a maximum error of $1.90$ m and $80\%$ of the error is $1$ m. When $E$ increases to $\sigma$ m, the accuracy becomes slightly worse with the maximum error being $2.90$ m and $80\%$ error being around $1.50$ m. As shown in Table \ref{table:AverageErr}, all of the average errors of SRL-KNN are around $\sigma/2$, which indicates that SRL-KNN is robust to localization error of the previous position. If the value of error $E$ is larger than $\sigma$, i.e., $E=3\sigma/2$ or $E=2\sigma$, the performance will degrade and the accumulated errors become more significant. The theoretical explanation is as follows. SRL-KNN implements a penalty function based on the previous location to discriminate the ambiguous locations. A location $\bm{l}_{j}$ is defined as an ambiguous point of $\bm{l}_{i}$ if their physical distance is larger than the grid size but their two vectors $\bm{f}_{i}$ and $\bm{f}_{j}$ have a fairly high Pearson correlation coefficient above the correlation threshold. We choose the value of the correlation threshold equal to the average correlation coefficients between $\bm{l}_{i}$ and all of its physical nearest neighbours, i.e., approximately $0.85$ in our database. Then all non-nearest-neighbour locations whose correlation coefficient above this threshold are considered as ambiguous points. Note that two locations are defined as physical neighbours if the physical distance between them within the grid size. The ambiguous distance $d_a$ is defined as the physical distance between a location and its ambiguous point. Pearson correlation coefficient $\rho(\bm{f}_{i},\bm{f}_{j})$  between $\bm{f}_{i}$  and $\bm{f}_{j}$ can be calculated as follows  
\[
\rho(\bm{f}_{i},\bm{f}_{j}) = \frac{1}{N-1} \sum_{n=1}^{N} (\frac{F^{i}_{n}-\mu_i}{\delta_i}) (\frac{F^{j}_{n}-\mu_j}{\delta_j}) 
\]
where $N$ is the number of available fingerprints, $\mu_i$, $\mu_j$ are the means of $\bm{f}_{i}$ and $\bm{f}_{j}$ respectively, $\delta_i$, $\delta_j$ are the standard deviations of $\bm{f}_{i}$ and $\bm{f}_{j}$ respectively. Fig. \ref{fig:fin_combine}(c) shows that if the error of previous location is within $d_a - \sigma$, the penalty function can provide higher likelihood to the potential locations near the correct current position and lower likelihood to the other ambiguous locations. Therefore, the estimation accuracy of the current location will not be adversely affected. In order to estimate $d_a$, Fig. \ref{fig:CDF}(d) illustrates the maximum and average ambiguous distances of all $365$ locations in the database. The average ambiguous distance $\bar{d_a}$ is around $4$ m ($2\sigma$) and the maximum value $d_a^{max}$ is above $12$ m ($6\sigma$). These results affirm that if the error of the previous locations is within $\bar{d_a} - \sigma = \sigma$, SRL-KNN is robust to the localization error of the previous position. Furthermore, according to the survey in \cite{Wu2018}, the percentage of stationary time can exceed $80\%$ for most mobile users. During the no movement period, the number of RSSI readings collected in one-location ($S_{2}$) is sufficient to improve the conventional KNN accuracy. Therefore, in order to enhance the accuracy when locating a user's position in a long trajectory, we can employ these stationary locations as aligning points where the prior locations can be ignored. In that case, some classical KNN approaches including RADAR \cite{Bahl2000}, WKNN \cite{Brunato2005} or STI-WKNN \cite{Zou2017} can be exploited to estimate the user's location.

In order to prove the consistent effectiveness of SRL-KNN, our algorithm is implemented with another published dataset, namely UJIIndoorLoc \cite{Torres-Sospedra2014}. The reported average localization error in \cite{Torres-Sospedra2014} is $7.9$ m. The training and validation data in all $3$ buildings of the databse from $2$ random phone users (Phone Id: $13$, $14$) are used to implement SRL-KNN algorithm. The maximum distance between $2$ consecutive locations in the testing trajectory can be up to $20$ m so $\sigma = 20$ m is chosen. Note that the grid size of UJIIndoorLoc is different from our collected database so the average localization error for UJIIndoorLoc is different from that reported previously. However, the relative accuracy comparison between SRL-KNN and conventional KNN, e.g., RADAR \cite{Bahl2000} or STI-WKNN \cite{Zou2017} can still reflect well the effectiveness of our algorithm. Table \ref{table:AverageErr2} shows the average errors in meter of SRL-KNN, RADAR, STI-WKNN for each separate building and for all $3$ buildings in general. These results consistently illustrate that SRL-KNN is more robust than other conventional KNN algorithms including RADAR \cite{Bahl2000} and STI-WKNN \cite{Zou2017}. For all $3$ buildings, the average error of SRL-KNN using mean fingerprint is $5.0$ m while the result of RADAR is $7.7$ m and STI-WKNN is $7.0$ m. Furthermore, Fig. \ref{fig:Ujdata} compares the CDF of localization errors between $3$ methods. In total, a $16$ m maximum localization error is recorded for SRL-KNN, $22$ m for STI-WKNN and the largest maximum localization error of $25$ m for RADAR. Besides, $80\%$ of the error is below $7$ m in the case of SRL-KNN, which is much lower than $13$ m and $12$ m in the case of RADAR and STI-WKNN, respectively.   
\section{Conclusions} \label{sec:conclude}
In conclusion, we have proposed a low complexity soft range limited KNN (SRL-KNN) for WiFi indoor localization. This algorithm exploits the information of previous positions and simultaneously applies the soft range limiting factor for fingerprint distance calculation to achieve more accurate and stable positioning performance. We demonstrated that SRL-KNN can address effectively some main challenges of KNN including the spatial ambiguity, RSSI instability and the RSSI short collecting time, especially when RSSI histogram is taken into account in calculating fingerprint distance. Experimental results have shown that SRL-KNN achieves the best accuracy of $0.66$ m with $80\%$ of the error within $0.89$ m, which outperforms existing KNN methods. In future research, we will apply the idea of the soft range limiting factor to other methods such as probabilistic methods or SVM to improve their performance.\\ 

\bibliographystyle{IEEEtran}
\bibliography{SRL_KNN}

\end{document}